\documentclass[conference]{IEEEtran}
\IEEEoverridecommandlockouts
\usepackage{amsmath,amssymb,amsfonts}
\usepackage{listings}
\usepackage{algorithm}
\usepackage{algorithmicx}
\usepackage[noend]{algpseudocode}
\usepackage{graphicx}
\usepackage{textcomp}
\usepackage{xcolor}
\usepackage{svg}
\usepackage{tikz}
\usetikzlibrary{shapes,positioning,calc}
\usepackage[hyphens]{url}
\usepackage[unicode, hidelinks]{hyperref}
\usepackage[hyphenbreaks]{breakurl}
\usepackage[style=ieee, bibencoding=utf8, citestyle=numeric-comp]{biblatex}
\def\BibTeX{{\rm B\kern-.05em{\sc i\kern-.025em b}\kern-.08em
    T\kern-.1667em\lower.7ex\hbox{E}\kern-.125emX}}
\usepackage{booktabs}
\usepackage{colortbl}
\usepackage{multirow}
\usepackage{flushend}
\usepackage[inline]{enumitem}
\usepackage{eqparbox}
\newdimen{\algindent}
\algnewcommand\LeftComment[2]{%
\hspace{#1\algindent}$\triangleright$ \eqparbox{COMMENT}{#2} \hfill %
}
\algnewcommand{\algorithmicand}{\texttt{ and }}
\algnewcommand{\algorithmicor}{\texttt{ or }}
\algnewcommand{\OR}{\algorithmicor}
\algnewcommand{\AND}{\algorithmicand}

\definecolor{mygreen}{rgb}{0,0.6,0}
\definecolor{mygray}{rgb}{0.5,0.5,0.5}
\definecolor{mymauve}{rgb}{0.58,0,0.82}

\begin{document}

\title{Numeric Truncation Security Predicate
}

\author{
\IEEEauthorblockN{
  Timofey Mezhuev\IEEEauthorrefmark{1}\IEEEauthorrefmark{2},
  Ilay Kobrin\IEEEauthorrefmark{1}\IEEEauthorrefmark{2},
  Alexey Vishnyakov\IEEEauthorrefmark{3}, and
  Daniil Kuts\IEEEauthorrefmark{1}
}
\IEEEauthorblockA{
  \IEEEauthorrefmark{1}Ivannikov Institute for System Programming of the RAS
}
\IEEEauthorblockA{
  \IEEEauthorrefmark{2}Lomonosov Moscow State University
}
\IEEEauthorblockA{
  \IEEEauthorrefmark{3}Yandex Cloud
}
Moscow, Russia \\
mezhuevtp@ispras.ru
kobrineli@ispras.ru
pmvishnya@gmail.com
kutz@ispras.ru
}

\maketitle

\begin{tikzpicture}[remember picture, overlay]
\node at ($(current page.south) + (0,0.65in)$) {
\begin{minipage}{\textwidth} \footnotesize
T. Mezhuev, I. Kobrin, A. Vishnyakov and D. Kuts, "Numeric Truncation Security Predicate,"
2023 Ivannikov ISPRAS Open Conference (ISPRAS), Moscow, Russian Federation, 2023,
pp. 84-92, DOI: 10.1109/ISPRAS60948.2023.10508174.\\
\copyright~2023 IEEE. Personal use of this material is permitted. Permission    
from IEEE must be obtained for all other uses, in any current or future media,  
including reprinting/republishing this material for advertising or promotional  
purposes, creating new collective works, for resale or redistribution to        
servers or lists, or reuse of any copyrighted component of this work in other   
works.
\end{minipage}
};
\end{tikzpicture}

\begin{abstract}
Numeric truncation is a widely spread error in software written in languages
with static data typing, such as C/C++ or Java. It occurs when the significant
bits of the value with a bigger type size
are truncated during value conversion to the smaller type.
Utilizing one of the most powerful methods for path exploration and automated
bug detection called dynamic symbolic execution (DSE), we propose the
symbolic security predicate for numeric truncation error detection, developed on
top of DSE tool Sydr. Firstly, we execute
the program on the data, which does not lead to any errors. During program execution
we update symbolic shadow stack and shadow registers to track symbolic sizes
of the symbolic variables to avoid false positives. Then, if we meet the instruction,
which truncates the symbolic variable, we build the security predicate, try to solve
it with the SMT-solver and in case of success save new input file to reproduce the error.
We tested our approach on Juliet Dynamic test suite for CWE-197 and achieved 100\%
accuracy. We approved the workability of our approach by detecting 12 new errors of
numeric truncation in 5 different real-world open source projects within
OSS-Sydr-Fuzz project. All of the errors were reported, most of the reports were
equipped with appropriate fixes, successfully confirmed and applied by project maintainers.

\end{abstract}

\begin{IEEEkeywords}
numeric truncation, security predicate, symbolic execution, concolic execution,
bug, weakness, dynamic analysis, automatic bug detection, binary analysis,
Juliet, error, sanitizer, DSE, SMT, SDL, CWE
\end{IEEEkeywords}

\section{Introduction}

Modern software is being rapidly developed, bringing new errors and
vulnerabilities~\cite{cwe}. Lots of companies integrate security
development lifecycle (SDL) into their workflow processes to detect errors as soon as
possible and prevent leaking them into the production~\cite{howard06}.

One of the most popular SDL technology is dynamic symbolic
execution~\cite{king76}, which produces new testcases by solving equations over symbolic
formulas that reflect dependencies on the input data. With symbolic execution
we can detect errors by constructing formulas corresponding to error conditions,
which we call security predicates~\cite{vishnyakov21},
and solving them, generating the input file to reproduce the error in case of success.

In this paper we propose the approach for detecting errors of numeric
truncation~\cite{cwe-197}. Numeric truncation error occurs when a value with the
bigger type size is
converted to the smaller type, which leads to truncation of value's significant
bits. This error is common for any programming language with static typing,
such as C/C++, Java, etc. Numeric truncation error can lead to incorrect program
execution or even to some vulnerabilities. For example, truncated value of allocated
memory size may later cause memory corruptions; or some security checks can be avoided due
to higher bits truncation of the tested value. 

To detect numeric truncation error we build the predicate
for every analyzed instruction, which moves only the part of the symbolic
value, discarding significant higher bits. Then we construct conjunction of the
predicate with the path predicate and pass the resulting formula to
SMT-solver~\cite{demoura08, niemetz23}. If the formula is satisfiable, we save
the generated input to be able to reproduce the error.

Consider the example of the C code on Listing~\ref{lst:simple-numtrunc-c} and
the corresponding assembly code on Listing~\ref{lst:simple-numtrunc-asm}, where
numeric truncation occurs. Firstly, we read the 4-byte value of type
\texttt{int32\_t} to the variable \texttt{a}. Then we move it to the 2-byte variable
\texttt{b} and print it. If the value of \texttt{a} is greater than
\texttt{INT16\_MAX} or lower than \texttt{INT16\_MIN}, numeric truncation will occur.

\begin{lstlisting}[language=C, basicstyle=\small\ttfamily, numbers=left,
                   caption={Simple numeric truncation example (C).},,
                   xleftmargin=3em, captionpos=b,
                   label=lst:simple-numtrunc-c]
int
main() {
    int32_t a = 0;
    scanf("%d", &a);
    int16_t b = a;
    printf("%d\n", b);
    return 0;
}
\end{lstlisting}

In the 3 line of
Listing~\ref{lst:simple-numtrunc-asm}
we see that the symbolic 4-byte value is loaded to \texttt{eax} register, and
then in the 4 line the lower 2 bytes of \texttt{eax} register are stored in
memory. When we observe such cases, we build the predicate and check it with
SMT-solver.

\begin{lstlisting}[language=C, basicstyle=\small\ttfamily, numbers=left,
                   caption={Simple numeric truncation example (assembly language).},
                   xleftmargin=3em, captionpos=b,
                   label=lst:simple-numtrunc-asm]
...
call   1030 <__isoc99_scanf@plt>
mov    eax,DWORD PTR [rbp-0x8]
(*\bfseries mov    WORD PTR [rbp-0xa],ax*)
\end{lstlisting}

This paper makes the following contributions:
\begin{itemize}
    \item We introduce security predicate for widely spread numeric truncation
        error, frequently met in the languages with static typing such as C/C++ or
        Java.
    \item We discovered 12 numeric truncation errors in 5 different real-world
    open source projects. We reported all the errors and most of them were successfully
    fixed.
\end{itemize}

The rest of the paper is organized as follows. Section~\ref{sec:related-work}
discusses some previously introduced methods for automated searching for integer errors
in binary code. Section~\ref{sec:numtrunc} introduces numeric truncation predicate
in more details and considers the motivating example of frequently met case, which enlightens
shadow stack and shadow registers structures importance in avoiding false positives.
Section~\ref{sec:shadow-stack} dives into the details of constructing and filling
shadow stack. Section~\ref{sec:shadow-regs} clarifies how shadow registers are updated.
In Section~\ref{sec:numtrunc-checkers} two main scenarios of checking for numeric
truncation error are discussed. Section~\ref{sec:implementation} reveals some details
of the proposed approach implementation. Section~\ref{sec:evaluation} is dedicated to
the approach testing on the Juliet Dynamic~\cite{juliet-dynamic} and its application to
real-world open source projects, giving examples of the most interesting errors found.
Section~\ref{sec:conclusion} concludes the paper.

\section{Related work}
\label{sec:related-work}

\subsection{Sydr security predicates}

Our method is based on the previously introduced security
predicates~\cite{vishnyakov21}, which were developed on the top of
Sydr~\cite{vishnyakov20}. Sydr is a tool for dynamic symbolic execution (DSE), 
that explores new execution paths in binary programs and enables error
detection. Sydr allows to search for integer overflow, null pointer dereference,
division by zero and out of bounds access errors with its security predicates.
Sydr analyzes every symbolic instruction of the binary file along the path given by
specific input data. Every time Sydr meets the instruction, execution of which may
potentially lead to an error, it constructs the corresponding security predicate, which is a
system of equations and inequalities over symbolic variables, then conjuncts it with
current path predicate and passes the resulting system to the SMT-solver. If the
system is satisfiable, Sydr prints the warning and saves the input for
reproducing the error.

\subsection{BRICK}

BRICK is a tool that is designed to work with binary files and search different
integer errors at runtime~\cite{chen09}. It searches for integer overflow, integer
underflow, signedness errors and assignment truncation. To search errors it uses
Valgrind as the core of its implementation, which helps to record some type
and value information of variables at runtime. It converts the binary code to
VEX intermediate representation and records necessary information. For
assignment truncation errors it searches for the loading and storing operations,
checking whether the value that is loaded or stored falls within the range of
its type.


\subsection{RICH}

Brumley et al.~\cite{brumley07} introduce RICH~-- a tool for C programs
instrumentation, which helps to prevent different integer errors, including integer
truncation errors. In their work the authors declare safe integer operations semantics.
RICH accepts C code as an input, compiles it to GCC intermediate representation and
then replaces every unsafe integer operation with the safe one, 
inserting the corresponding checks and giving the warning if the
security check is violated. In fact, RICH is pretty similar to Google
Sanitizers~\cite{google_san}, which work with LLVM IR.


\subsection{IntFind}

Chen et al.~\cite{chen09_2} present IntFind~-- a tool for searching for integer
errors in binary code. It searches for integer overflow, integer underflow,
signedness errors and assignment truncation errors. Firstly, it gets information
about variables types with the help of the static analysis, based on decompiler. Then it
passes this information to dynamic analysis part of the tool, which consists of
taint analysis tool and error detection tool. Detection tool gets the information
about types from the static analysis step and checks for the errors only those
instructions, which were tainted after taint analysis step.


\subsection{IntScope}

IntScope is a tool for detecing integer overflow errors in binary code with the
help of symbolic execution~\cite{wang09}. It aims to check for potential overflow in
arithmetic instructions, when the potentially overflowed result reaches
sensitive places, like calls of \texttt{malloc} functions, memory accesses,
conditional jumps, etc. To make the check more accurate, it considers the numeric
value as unsigned if the error sink is a memory allocation function. Otherwise
it get signedness from previously met instructions of conditional jumps.

\section{Numeric Truncation Security Predicate}
\label{sec:numtrunc}

Numeric truncation security predicate is developed on top of Sydr~\cite{vishnyakov20}
symbolic engine and organized like other security predicates in Sydr~\cite{vishnyakov21}.
We utilize Triton~\cite{saudel15} framework for symbolic execution to construct formulas
for security predicates.

We analyze every symbolic instruction during the symbolic execution. When we meet instructions
like \texttt{mov}, \texttt{movsx}, \texttt{movzx}, \texttt{cbw}, \texttt{cwde}, \texttt{cdqe}, that may potentially truncate the
value, we check them for the numeric truncation error by building and solving the
correspoinding security predicate. If the initial size of the symbolic value, located
in register or memory, is bigger than the source operand size of aforementioned
\texttt{mov*} instructions, or the size of the symbolic value located in \texttt{rax}
register is bigger than the size of the value being extended after \texttt{cbw}, \texttt{cwde}, \texttt{cdqe}
instructions execution, then numeric truncation error may occur. For example, there
is a number with 4 significant bytes in \texttt{eax} register. Then we meet the instruction
\texttt{mov bx, ax} where only 2 bytes of \texttt{eax} register are taken and stored
in \texttt{bx} register. If 2 higher bytes of \texttt{eax} register contained some
significant value, then the numeric truncation error emerges.


For \texttt{mov}, \texttt{movsx}, \texttt{movzx} instructions the basic idea is to check whether the size
of the source operand is less than the size of the symbolic variable located in memory
or register, respectively. In all \texttt{mov}, \texttt{movsx}, \texttt{movzx} instructions the source
operand must correspond to $extract(high, low, \phi)$ formula, which extracts bits from
$low$ to $high$ of $\phi$ formula, to be able to lead to the numeric truncation error.
So to build the numeric truncation security predicate we search for aforementioned
instructions whose second operand corresponds to $extract$ formula. 
If we meet such instruction, we can use $high$ bound of the extraction, symbolic
variable formula $\phi$ and its size to check whether the bits from $high + 1$ to the
size of the variable can be significant.

Unfortunately, the $extract$ formula corresponding to the second operand of the
instruction is a necessary, but not sufficient condition for potential numeric truncation
error. Suppose that we have \texttt{movsx ecx, ax} instruction, where in \texttt{eax}
register we have 2-byte symbolic variable. If we check the formula of the source operand
for this case, we will get something like $extract(15, 0, \phi_{eax})$. The thing is that
registers in Triton framework are atomic and there are no subregisters as independent entities.
So if the symbolic value of any size is stored in register, the whole register is
considered as symbolic, and the formula corresponding to a subregister corresponds
to the $extract$ formula. 
Such cases may lead to redundant SMT-solver calls, that affect Sydr performance, and
even to false positive error reports.
To overcome this problem, we need a structure to store the sizes of the symbolic variables
located in registers.




When we meet \texttt{cbw}, \texttt{cwde}, \texttt{cdqe} instructions, we do the following:
\begin{enumerate}
  \item Calculate the register size that is being extended. For \texttt{cbw} it
  is 1 byte, for \texttt{cwde} it is 2 bytes and for \texttt{cdqe} it is 4 bytes.
  \item Check whether the size of the symbolic variable located in \texttt{rax} register
  is bigger than the size of the \texttt{rax} register part being extended
  (i.e.\ significant higher bits of the \texttt{rax} will be overwritten by constant
  zeroes or ones). If so, we build the security predicate for the cropped part of the
  \texttt{rax} register.
\end{enumerate}

We need to know the size of the symbolic variable located in \texttt{rax}
register to analyze \texttt{cbw}, \texttt{cwde}, \texttt{cdqe} instructions. We don't have this information
from Triton, because if there is a symbolic variable located in \texttt{rax} register,
the whole register is considered symbolic. This is the second reason why we need a structure
to track symbolic variable sizes located in registers.

To build the security predicate we need a size of the symbolic value being truncated,
a size of the resulting truncated value, and a formula corresponding to the symbolic
value being truncated. Let us call the size of the original symbolic
value \textit{high}, the size of the resulting truncated value \textit{low}, and
the formula of the symbolic value being truncated $\phi_{var}$; 
sizes are represented in bits.
Using these parameters we build the numeric truncation
security predicate for signed and unsigned cases.

Formulas in Triton are organized as abstract syntax trees (AST), where a single node
corresponds to an operation over the bitvectors. Formulas can be translated into a format
native for SMT-solver.

Let $\phi_{trunc} = extract(high, low, \phi_{var})$~-- significant bits being truncated
from the original symbolic value. $sz$~-- size of $\phi_{trunc}$ formula, i.e.
$sz = high - low + 1$. $bv(0, sz)$~-- bitvector of $sz$ zeros and $bv(1, sz)$~-- 
bitvector of $sz$ ones. Then we can derive formulas for signed and unsigned numeric
truncation errors:
\begin{enumerate}
  \item Signed numeric truncation predicate: \\
  $not((\phi_{trunc} == bv(0, sz)) \lor (\phi_{trunc} == bv(1, sz)))$
  \item Unsigned numeric truncation predicate: \\
  $not(\phi_{trunc} == bv(0, sz))$
\end{enumerate}


Formula for signed truncation means that if there are not only zeroes or ones in
the cropped part of the symbolic value then this is a numeric truncation.
For unsigned case numeric truncation occurs if there is something except of zeroes
in the cropped part.

We get the information about the symbolic variable signedness in the same way
it is obtained for the integer overflow security predicate, which was presented by
Vishnyakov et al.~\cite{vishnyakov21}. 
The signedness of the symbolic variable is derived in 2 ways. Firstly, we check
whether the value was read with functions like \texttt{fscanf}, \texttt{scanf},
etc., that use \texttt{strto*} functions inside. If so, we get the signedness from
the function been used: signed for \texttt{strtol*} case, unsigned for
\texttt{strtoul*} case. Secondly, if \texttt{strto*} functions weren't used,
we get the signedness with the backward slicing algorithm. We start to iterate from
the last branch in path predicate and search for the first branch that can provide 
information about signedness. For example, \texttt{jl} branch instruction tells us
that the value is signed, while \texttt{ja} branch instruction tells us that the value
is unsigned.

After signedness detection and predicate construction, we conjunct the security
predicate with the current path predicate and pass the resulting predicate
to SMT-solver. If the predicate is satisfiable, we print the warning and
save the input file generated by SMT-solver for later error reproducing.


In addition to the problem with tracking the size of the symbolic value located in
registers described above, there are some other cases which may lead to false
positives and which we need to take into account.
We need to get rid of false positives as much as possible, because they make
the whole analysis less productive.

Let's consider a prime example of the false positive case on the
Listing~\ref{lst:false-positive-c}.
\begin{lstlisting}[language=C, basicstyle=\small\ttfamily, numbers=left,
                   caption={Numeric Truncation (Linux x32).},
                   xleftmargin=3em, captionpos=b,
                   label=lst:false-positive-c]
void
foo(int16_t a, int8_t b)
{
    printf("%d%d", a, b);
}

int
main()
{
    int32_t int32Primitive;
    scanf("%d", &int32Primitive);
    foo(int32Primitive, int32Primitive);
    return 0;
}
\end{lstlisting}

Numeric truncation in this example occurs when we put \texttt{int32\_t} variable to function
parameters that have types \texttt{int16\_t} and \texttt{int8\_t}, respectively.

Then let's compile this code with \textit{gcc -m32 prog.c -O0 -o prog} and have a
look at its significant part of Assembly code on Listing~\ref{lst:false-positive-asm}.
\begin{lstlisting}[language={C}, numbers=left, captionpos=b,
                   basicstyle=\small\ttfamily, xleftmargin=3em,
                   caption={Numeric Truncation (Assembly language).},
                   label=lst:false-positive-asm]
00001255 <main>:
call   10c0 <__isoc99_scanf@plt>
add    esp,0x10
mov    eax,DWORD PTR [ebp-0x10]
(*\bfseries movsx  edx,al*)
mov    eax,DWORD PTR [ebp-0x10]
(*\bfseries cwde*)
push   edx
push   eax
call   120d <foo>
...
0000120d <foo>:
...
mov    edx,DWORD PTR [ebp+0x8]
mov    ecx,DWORD PTR [ebp+0xc]
(*\bfseries mov    WORD PTR [ebp-0xc],dx*)
mov    edx,ecx
(*\bfseries mov    BYTE PTR [ebp-0x10],dl*)
\end{lstlisting}

Here we can see that on binary level numeric truncation takes place at lines 5 and 7.
So the actual sizes of the symbolic values located in \texttt{edx} and \texttt{eax}
are 1 byte and 2 bytes respectively.
Then they are pushed as arguments for \texttt{foo} function and there they are popped into \texttt{edx} and
\texttt{ecx} registers. In lines 16 and 18 they are converted again to 1 and 2 bytes, so
here the false positive truncation may happen, because the formulas of the operands are
$extract$ and we may consider this case as a potential numeric truncation. Even if the false positive doesn't
occur, there will be an extra query to the SMT-solver, which might spend solving time and thus decrease Sydr
effectiveness.
That is why we need to somehow store the actual sizes of symbolic values located
on stack and registers and to uphold these sizes to be relevant. We overcome
this problem by introducing shadow stack map for storing actual sizes of symbolic
values located in memory, and shadow registers map for storing actual sizes of
symbolic values located in registers.

If we head back to the example presented on Listing~\ref{lst:false-positive-c} and
Listing~\ref{lst:false-positive-asm}, we can clearly see the
sense of these shadow structures. After the instruction at line 5 we can see that
\texttt{edx} stores only 1 significant byte, the same is for \texttt{eax} at
line 7: only 2 significant bytes are stored there. Before entering \texttt{foo}
function \texttt{edx} and \texttt{eax} registers are pushed. If we store their
actual symbolic sizes using shadow stack, we can avoid false positive cases and
redundant calls to SMT-solver at lines 16 and 18.
We need to update actual size of memory/registers when we process such instructions
as \texttt{mov/movsx/movzx}, \texttt{pop/push} and conversion instructions
\texttt{cbw}, \texttt{cwde}, \texttt{cdqe}.



\begin{figure*}[t]
\caption{Numeric Truncation Predicate scheme}
\center
\includegraphics[scale=0.6]{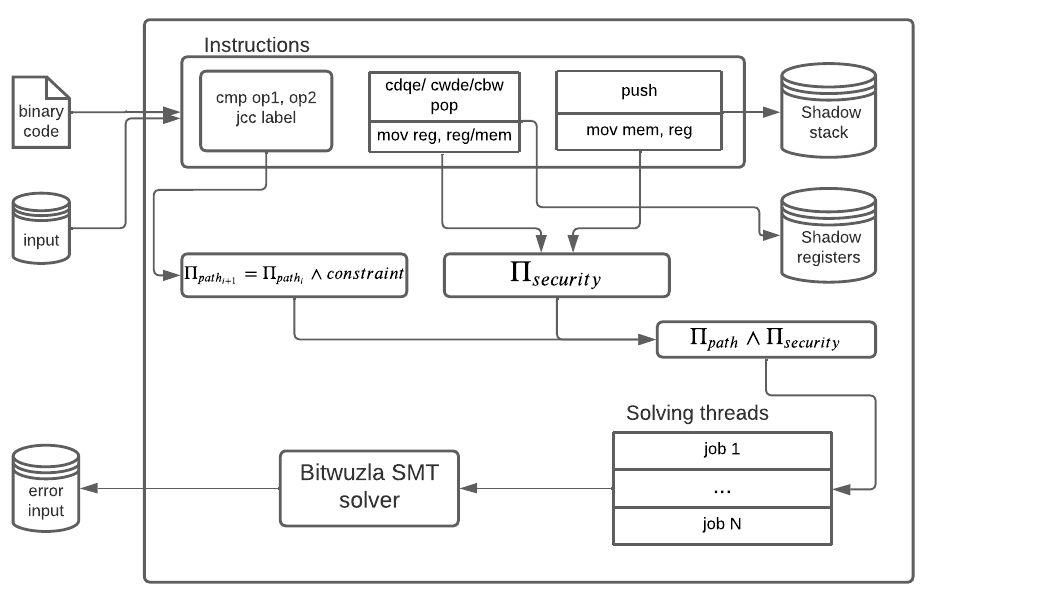}
\label{fig:numtrunc-scheme}
\end{figure*}

\section{Symbolic Shadow Stack}
\label{sec:shadow-stack}

To fill the shadow stack we analyze \texttt{mov}, \texttt{movsx}, \texttt{movzx}, \texttt{push} instructions.
\begin{enumerate}
  \item For \texttt{mov}, \texttt{movsx}, \texttt{movzx} instructions we check that a destination 
  operand is memory and source operand is symbolic. Thus, the second operand
  is a register in this case. If the register is in shadow registers map, then we
  fill the shadow stack for the address corresponding to the destination operand
  with the minimum of the size from shadow registers map and the size of source operand
  (we take minimum to avoid errors in shadow registers map). If the register is not in the
  shadow register map, then we just consider the size of the value located on stack equal to the size of a source operand.
  \item For \texttt{push} instruction we fill shadow stack for current \texttt{sp} value. 
  If the operand is symbolic, then we can update shadow stack with the minimum of
  the size from the shadow registers map (if there is an occurrence) and the size
  of source operand. If the operand register is not presented in shadow registers
  map, we fill the shadow stack with the size of the source operand.
\end{enumerate}

Our shadow stack is organized as a stack of maps, where we map from stack address to symbolic
size of variable located at this address for all frames in current callstack. When a function
is called, a new frame is pushed; after returning from the function this frame is popped from
the shadow stack. To find an address in this stack we need to find a current frame and then search
for this address in the map. Thus, we prevent memory consumption, because our shadow stack always
corresponds only to those functions which are currently in the callstack.

\section{Symbolic Shadow Registers}
\label{sec:shadow-regs}
To fill the shadow registers map we analyze \texttt{mov}, \texttt{movsx}, 
\texttt{movzx}, \texttt{cbw}, \texttt{cwde}, \texttt{cdqe}
and \texttt{pop} instructions.
\begin{enumerate}
  \item For \texttt{mov}, \texttt{movsx}, \texttt{movzx} instructions the destination operand must be register
  and source operand must be symbolic. If a source operand is memory, then we search for its address
  in the shadow stack. If it was found, then the shadow register size is equal to a minimum of
  the size from shadow stack and the size of the source operand (we take minimum to avoid errors
  in shadow stack). If address is not in the shadow stack, then we fill the shadow register with
  the size of the source operand. In case when the source operand is a register we use the same
  algorithm that is presented above, but actual size is searched in the shadow registers map.
  \item In case of \texttt{pop} instruction we try to get symbolic size from shadow stack for
  current \texttt{sp} value and then save it in shadow registers map for the register operand.
  \item For \texttt{cbw}, \texttt{cwde}, \texttt{cdqe} instructions we calculate register size that is
  being extended. It will be 1, 2, 4 bytes respectively. So the actual register size becomes
  equal to the extendable register size. For example, if we analyze \texttt{cbw} instruction,
  the size of the \texttt{rax} register in the shadow registers map will be 1 byte.
  \item In case of other instructions we get all the registers written by the instruction and
  update the shadow registers map for these registers with their new symbolic sizes. This
  is needed to keep up-to-date actual symbolic sizes when the arithmetic operations are performed.
  For example, \texttt{eax} register has 1 significant byte. Then \texttt{add eax, 0xffffff00}
  instruction is executed and after that all 4 bytes of \texttt{eax} become significant.
  When we use this written register implementation, we can easily set the right size of \texttt{eax}
  register to 4 bytes. After that, we can check \texttt{eax} register for numeric truncation
  correctly.
\end{enumerate}

There is also an issue with \texttt{rax/eax} register. To speed up symbolic execution
and make Sydr more effective, we build semantics for some standard library functions~\cite{vishnyakov21}
(e.g. \texttt{strtoul}, \texttt{malloc}, \texttt{strcpy}, etc.).
Thus, we skip symbolic execution of these functions bodies and just update symbolic states
for all registers and memory regions directly and indirectly affected by them.
However, skipping symbolic execution of these functions bodies leads us to missing some
instructions which write to the \texttt{rax} register. Thus, the occurrence of \texttt{rax}
register in shadow registers map can become incorrect.
To ovecome this problem every time we build function semantics we check whether this
function returns a value, and if the \texttt{rax} register is symbolic, we update
its occurrence in shadow registers map. Otherwise we remove it.

\section{Numeric Truncation Checkers}
\label{sec:numtrunc-checkers}

We highlight three numeric truncation checking scenarios.

The first one is the analyzing \texttt{mov}, \texttt{movsx}, \texttt{movzx} instructions, when
the source operand is memory. In this case we search the shadow stack for
operand memory address. If this address is found, we check whether the size of
the source operand is less than the actual size of the symbolic variable from the
shadow stack. If so, we check for numeric truncation error in the cropped part
of memory and build numeric truncation security predicate as described in Section~\ref{sec:numtrunc}.

The second scenario of checking for the numeric truncation error is the analyzing
\texttt{mov}, \texttt{movsx}, \texttt{movzx} instructions when the source operand is register.
\begin{enumerate}
  \item Firstly, we search in shadow registers map for the source operand register.
  If it is presented, we check whether the size of the source operand is less
  than the actual size of symbolic value located in the register, and build
  security predicate.
  \item If the register is not presented in shadow registers map, we build predicate
  only if the source operand is a subregister and it corresponds to an $extract$ formula
  (see Section~\ref{sec:numtrunc}).
\end{enumerate}

The last scenario is analyzing conversion instructions
like \texttt{cbw}, \texttt{cwde}, \texttt{cdqe}.
\begin{enumerate}
  \item If \texttt{rax} register is not presented in shadow registers map, then we
  don't build numeric truncation security predicate (we can't say for sure that this
  case won't be false positive).
  \item If \texttt{rax} register is presented in shadow registers map, we calculate
  the size that is being extended. It will be equal to 1, 2 or 4 bytes respectively.
  \item If the size obtained from shadow registers map is greater than the part of register
  being extended, we build the security predicate.
\end{enumerate}

\section{Implementation}
\label{sec:implementation}

Numeric truncation security predicate is developed on top of
Sydr~\cite{vishnyakov20} tool for dynamic symbolic execution, which is based on
Triton~\cite{saudel15} symbolic execution framework and dynamic binary
instrumentation tool DynamoRIO~\cite{bruening04}.
Our approach uses Bitwuzla~\cite{niemetz23} SMT-solver for solving formulas and
generating inputs.
Numeric truncation security
predicate checker is written in C++ and takes about 1500 lines of code. For now the
implementation supports only x86 (32 and 64 bit) architecture, but later it may
be implemented for ARM architecture.

Figure~\ref{fig:numtrunc-scheme} represents the implementation of numeric truncation
security predicate in Sydr. The scheme also describes the general implementation of all
security predicates in Sydr except for the steps of updating shadow stack, shadow registers and
the type of target instructions. We analyze instructions on the concrete path provided by
the input file. The path predicate is updated on each conditional jump.
We build the security predicate when we meet certain instructions.
Then we conjunct the security predicate
with the path predicate and push a job for the solver to the queue. Solving threads will later
asynchronously pick up the job and pass the predicate to Bitwuzla SMT-solver, which
will generate the error-revealing input in case of predicate's satisfiability.

\section{Evaluation}
\label{sec:evaluation}


\subsection{Juliet Dynamic}

Juliet Dynamic was created on top of Juliet test suite~\cite{juliet} and was presented
by Vishnyakov et al.~\cite{vishnyakov21}. It tests the
instrument in the following way. Firstly, it builds all the tests for specified CWE
to binaries in 32-bit and 64-bit modes. Then it runs the tool under test on all the
binaries with the sample input data. If the tool warns about the error in the test,
where the error was located, the result is true positive. If the error was not located in
this test, the result is false positive. If there is no warning while the error is located
in the test, the result is false negative, and at last if there is no warning and no
error was located in the test, the result is true negative. After that all true
positive results are being verified with binaries built with sanitizers. If the error
can be reproduced with the generated input and binary built with sanitizers, then it
is still true positive. Otherwise it is false negative. After that Juliet Dynamic
computes true positive rate, true negative rate and accuracy.

We tested our numeric truncation security predicate on Juliet Dynamic~\cite{juliet-dynamic}
and our approach showed 100\% accuracy result without sanitizers verification~\cite{ubsan}.
Unfortunately, we were unable to properly test our approach with sanitizers verification.
The thing is that Undefined Behavior sanitizer considers numeric truncation errors
only those with the implicit type conversion, while explicit type conversion isn't
considered to be an error. On the other hand, all of the tests in Juliet~\cite{juliet}
and thus in Juliet Dynamic~\cite{juliet-dynamic} for CWE-197, which corresponds to
numeric truncation error, are made with an explicit numeric truncation.
So regardless of the input data all the tests are considered to have no error from the
sanitizers point of view.

\subsection{Open Source Trophies}
Our tool has found 12 new numeric truncation errors in open-source projects. Error numbers
for every project are showed in Table~\ref{tbl:error-numbers}.
All of them were reported and fixed, and most of them were approved by projects maintainers.

\begin{table}[h]
\caption{Errors in Open-Source Projects}
\label{tbl:error-numbers}
\begin{center}
\scriptsize
\begin{tabular}{ c|c }
\toprule
\textbf{Project} & \textbf{Detected errors number} \\
  \hline
  nDPI~\cite{nDPI} & 7~\cite{nDPI-err-reader, nDPI-err-tls, nDPI-err-main, nDPI-err-analyze, nDPI-err-kerberos, nDPI-err-rtcp, nDPI-err-diameter} \\
  \hline
  libpcap~\cite{libpcap} & 2~\cite{libpcap-err-util, libpcap-err-usb} \\
  \hline
  FreeImage~\cite{freeimage} & 1~\cite{freeimage-err} \\
  \hline
  LibTIFF~\cite{libTIFF} & 1~\cite{libTIFF-err} \\
  \hline
  unbound~\cite{unbound} & 1~\cite{unbound-err} \\
\bottomrule
\end{tabular}
\end{center}
\end{table}

\subsection{nDPI}
nDPI is an open source library for deep-packet inspection~\cite{nDPI}.
We have found 7 numeric truncation errors in nDPI project and here we will show
2 most interesting of them.

On Listing~\ref{lst:ndpi-error-1} the variable \texttt{com\_code} has type
\texttt{u\_int16\_t}, but the value to the right of the assignment at line 12
has type \texttt{int} due to integer promotion, so here
a numeric truncation error may occur. Then \texttt{com\_code} variable is used in
\textit{if} at line 16. If this variable is truncated to one of those constants,
the function will perform the wrong behavior.

Firstly, our approach found this error and generated an input that simply leads to a numeric truncation at
line 12 on Listing~\ref{lst:ndpi-error-1}, but the value wasn't truncated to one of
the constants enumerated in the code, so the presented function just returned the error.
For this example we patched our approach to not just find the numeric truncation error,
but to truncate the value in such way that it will match one of these constants.
Our approach managed to generate such input and thus we passed this check with a very big
value before the truncation.

We fixed this error by changing type of the \texttt{com\_code} variable to
\texttt{uint32\_t}~\cite{nDPI-err-diameter}.

\begin{lstlisting}[language=C, basicstyle=\small\ttfamily, numbers=left,
                   caption={Numeric Truncation in nDPI.},
                   xleftmargin=3em, captionpos=b,
                   label=lst:ndpi-error-1]
typedef enum {
  AC = 271,
  AS = 274,
  CC = 272,
  CE = 257,
  DW = 280,
  DP = 282,
  RA = 258,
  ST = 275
} com_type_t;
...
(*\bfseries u\_int16\_t com\_code *)= diameter->com_code[2]
+ (diameter->com_code[1] << 8)
+ (diameter->com_code[0] << 8);

if(com_code == AC || com_code == AS ||
  com_code == CC || com_code == CE ||
  com_code == DW || com_code == DP ||
  com_code == RA || com_code == ST)
  return 0;

...
\end{lstlisting}

On Listing~\ref{lst:ndpi-error-2} a numeric truncation error can occur at several lines.
In \texttt{struct ndpi\_analyze\_struct} \texttt{min\_val, max\_val, sum\_total, *values}
all fields have \texttt{u\_int32\_t} type, while \texttt{value} variable, passed as a
function argument, has \texttt{u\_int64\_t} type, so a numeric truncation error may occur at
lines 8, 11, 13, 16 and 20.

Our approach found this error and generated input for its reproducing. We fixed it
by changing types of the aforementioned fields to \texttt{u\_int64\_t}~\cite{nDPI-err-analyze}.

\begin{lstlisting}[language=C, basicstyle=\small\ttfamily, numbers=left,
                   caption={Numeric Truncation in nDPI at several lines simultaneously.},
                   xleftmargin=3em, captionpos=b,
                   label=lst:ndpi-error-2]
void ndpi_data_add_value(
    struct ndpi_analyze_struct *s,
    const u_int64_t value) {
  if(!s)
    return;

  if(s->sum_total == 0)
    (*\bfseries s->min\_val *)= s->max_val = value;
  else {
    if(value < s->min_val)
      (*\bfseries s->min\_val *)= value;
    if(value > s->max_val)
      (*\bfseries s->max\_val *)= value;
  }

  (*\bfseries s->sum\_total *)+= value,
      s->num_data_entries++;

  if(s->num_values_array_len) {
    (*\bfseries s->values[s->next\_value\_insert\_index] *)
      = value;
  }

  ...
}
\end{lstlisting}

\subsection{LibTIFF}

LibTIFF is a library for reading and writing Tagged Image File Format
(abbreviated TIFF) files~\cite{libTIFF}.

On Listing~\ref{lst:libtiff-error} variable \texttt{m} has type \texttt{uint16\_t},
structure \texttt{TIFFDirEntry} has field \texttt{dir\_tag} with a type \texttt{uint16\_t}.
But on the right side of operator at line 23 there
is an integer type value (due to the integer promotion), so the numeric
truncation may occur. Our tool has found input when \texttt{o->tdir\_tag} is
equal to 65535 (\texttt{UINT16\_MAX}), so the value \texttt{o->tdir\_tag + 1} truncates
to zero in variable \texttt{m}. Then \texttt{m} is used in \textit{if} at line
14, and therefore \texttt{o->tdir\_tag < m} expression cannot be true after
truncation, so the break from the cycle won't occur even with not ascending
order of tags. This bug was fixed by changing the type \texttt{uint16\_t} of the variable
\texttt{m} to the type \texttt{uint32\_t}~\cite{libTIFF-err}.

\begin{lstlisting}[language=C, basicstyle=\small\ttfamily, numbers=left,
                   caption={Numeric Truncation in libTIFF.},
                   xleftmargin=3em, captionpos=b,
                   label=lst:libtiff-error]
static void TIFFReadDirectoryCheckOrder(
  TIFF *tif, TIFFDirEntry *dir,
  uint16_t dircount)
{
  static const char module[]
  = "TIFFReadDirectoryCheckOrder";
  uint16_t m;
  uint16_t n;
  TIFFDirEntry *o;
  m = 0;
  for (n = 0, o = dir; n < dircount; n++, 
       o++)
  {
      if (o->tdir_tag < m)
      {
          TIFFWarningExtR(
          tif, module,
          "Invalid TIFF directory;"
          "tags are not sorted in "
          "ascending order");
          break;
      }
      (*\bfseries m *)= o->tdir_tag + 1;
  }
}
\end{lstlisting}

\subsection{unbound}
Unbound is a validating, recursive, caching DNS resolver~\cite{unbound}.

The code fragment containing an error is presented on Listing~\ref{lst:unbound-error}. 
In function \texttt{sldns\_str2wire\_nsec\_buf}
at line 8 variable \texttt{t} has type \texttt{uint16\_t}, function
\texttt{sldns\_get\_rr\_type\_by\_name} returns enum type \texttt{sldns\_rr\_type}
that has \texttt{unsigned int} type value. The same is in \texttt{sldns\_str2wire\_type\_buf}
function in line 17. Enumeration \texttt{sldns\_rr\_type}
has values bigger than or equal to 65535. Our tool has found input
where \texttt{sldns\_get\_rr\_type\_by\_name} function returns \texttt{atoi(name + 4)} (line 27) and its value
occurs to be bigger than
65535, so the numeric truncation error occurs. We fixed this by adding additional checkers in
\texttt{sldns\_get\_rr\_type\_by\_name} function:
if \texttt{atoi(name + 4)} is bigger than \texttt{LDNS\_RR\_TYPE\_LAST} (it is the biggest element
in \texttt{sldns\_rr\_type} enumeration and it is less than \texttt{UINT16\_MAX}), then we return zero. In caller functions there was already
a check for a zero value returned from this function~\cite{unbound-err}.

\begin{lstlisting}[language=C, basicstyle=\footnotesize\ttfamily, numbers=left,
                   caption={Numeric Truncation in unbound.},
                   xleftmargin=2em, captionpos=b,
                   label=lst:unbound-error,]
int sldns_str2wire_nsec_buf(
  const char* str, uint8_t* rd, size_t* len)
{
...
while(sldns_buffer_remaining(&strbuf) > 0 &&
  sldns_bget_token(&strbuf, token, 
  delim, sizeof(token)) != -1) {
  (*\bfseries uint16\_t t *)= sldns_get_rr_type_by_name(token);
  ...
}
...
}

int sldns_str2wire_type_buf(const char* str,
  uint8_t* rd, size_t* len)
{
  (*\bfseries uint16\_t t *)= sldns_get_rr_type_by_name(str);
  ...
}

sldns_rr_type
sldns_get_rr_type_by_name(const char *name)
{
...
if (strlen(name) > 4 && strncasecmp(name,
    "TYPE", 4) == 0) {
  return atoi(name + 4);
}
...
}
\end{lstlisting}

\subsection{libpcap}
Libpcap provides a portable framework for low-level network monitoring~\cite{libpcap}.

On Listing~\ref{lst:libpcap-error} at line 4 variable \texttt{size} has type
\texttt{uint16\_t}, type of value on the right side of the assignment operator is \texttt{int} (due to integer promotion),
so the numeric truncation may occur. Our tool has found input where the value of
variable \texttt{size} truncates to zero. Then size is used in \textit{if} operators at lines
6 and 11, so the truncation may cause a wrong behavior. To fix this error we changed
type \texttt{uint16\_t} of the variable \texttt{size} to the type \texttt{u\_int}~\cite{libpcap-err-util}.

\begin{lstlisting}[language=C, basicstyle=\small\ttfamily, numbers=left,
                   caption={Numeric Truncation in libpcap.},
                   xleftmargin=3em, captionpos=b,
                   label=lst:libpcap-error]
...
size = tlv->tlv_length;
if (size % 4 != 0)
  (*\bfseries size *)+= 4 - size % 4;

if (size < sizeof(nflog_tlv_t)) {
  /* Yes. Give up now. */
  return;
}

if (caplen < size || length < size) {
  /* No. */
  return;
}
...
\end{lstlisting}

\subsection{freeimage}
FreeImage is an Open Source library project for developers who would like to
support popular graphics image formats like PNG, BMP, JPEG, TIFF and others
as needed by today's multimedia applications~\cite{freeimage}.

The part of FreeImage code with the error is presented on Listing~\ref{lst:freeimage-error}.
Type of variable \texttt{end} at line 2 is
\texttt{unsigned}, but \texttt{ftell} function returns \texttt{unsigned long}
value, so here the numeric truncation may occur. Then \texttt{end} is used in \textit{while}
cycle in line 14, so the wrong behavior may emerge. So here we suggested to change
the type \texttt{unsigned} of variable \texttt{end} to \texttt{unsigned long}
type~\cite{freeimage-err}.

\begin{lstlisting}[language=C, basicstyle=\small\ttfamily, numbers=left,
                   caption={Numeric Truncation in freeimage.},
                   xleftmargin=3em, captionpos=b,
                   label=lst:freeimage-error]
...
(*\bfseries end *)= ftell(ifp) + size;
if (!memcmp(tag, "RIFF", 4) ||
    !memcmp(tag, "LIST", 4))
{
  int maxloop = 1000;
  get4();
  while (ftell(ifp) + 7 < end &&
        !feof(ifp) && maxloop--)
    parse_riff(maxdepth-1);
}
else if (!memcmp(tag, "nctg", 4))
{
  while (ftell(ifp) + 7 < end)
  {
  if (feof(ifp))
    break;
    i = get2();
    size = get2();
    if ((i + 1) >> 1 == 10 && size == 20)
      get_timestamp(0);
    else
      fseek(ifp, size, SEEK_CUR);
  }
}
...
\end{lstlisting}

\section{Conclusion}
\label{sec:conclusion}

We propose security predicate for numeric truncation error detection in the
dynamic symbolic execution tool Sydr~\cite{vishnyakov20, vishnyakov21}. During
the target program execution we track and update the symbolic shadow stack and
shadow register structures, which help to avoid false positive cases. For each
instruction, which potentially leads to a numeric truncation, we build the
security predicate to check whether the higher bits of the truncated value can
contain significant value. If the security predicate is satisfiable, we print
the warning and save the input for error reproducing. We've tested numeric
truncation predicate
on the Juliet Dynamic~\cite{juliet-dynamic} and reached 100\% accuracy.
We've applied our approach to different real-world open source projects and found totally 12
errors in 5 different projects~\cite{nDPI-err-reader, nDPI-err-tls, nDPI-err-main, nDPI-err-analyze,
nDPI-err-kerberos, nDPI-err-rtcp, nDPI-err-diameter, libpcap-err-util, libpcap-err-usb,
freeimage-err, libTIFF-err, unbound-err}. All of the found errors were reported and most
of them were successfully fixed.

\printbibliography

\end{document}